\documentclass[aps,prl,twocolumn,nobibnotes,superscriptaddress,longbibliography]{revtex4}
\usepackage{graphicx}
\usepackage{dcolumn}
\usepackage{bm}
\usepackage{physics}
\usepackage{amsmath}
\usepackage{amssymb}
\usepackage{array}
\usepackage{soul}
\usepackage{color}
\usepackage{xcolor} 
\usepackage{float}
\newcolumntype{P}[1]{>{\centering\arraybackslash}p{#1}}
\usepackage[colorlinks=true, breaklinks=true, linkcolor=blue, citecolor=blue, urlcolor=blue]{hyperref}

\newcommand{\bonnpi}{Physikalisches Institut, University of Bonn, Nussallee 12, 53115 Bonn, Germany}
\newcommand{\pks}{Max Planck Institute for the Physics of Complex Systems, N\"othnitzer Str.~38, 01187 Dresden, Germany}
\newcommand{\cea}{Univ. Grenoble Alpes, CEA, Grenoble INP, IRIG, Pheliqs, F-38000 Grenoble, France}

\begin{document}

\title{Fluctuation-Induced Bistability in the Dissipative Dynamics of Generic  Cavity-Matter Quantum Systems
}
\date{\today}

\begin{abstract}
We demonstrate that fluctuation-induced bistability is a robust and generic phenomenon in strongly interacting many-body systems with strong light-matter coupling. 
We identify a common microscopic mechanism based on resonances between photonic transitions and many-body energy scales, unifying the emergence of fluctuation-induced bistability across a broad class of models, including interacting spins, fermions, and bosons coupled to cavity modes. 
We develop complementary methods to study both the steady-state properties of fluctuation-induced bistability and its dynamical formation at finite times. In particular, we introduce the dressed-state rate equation approach, which reveals rich metastable dynamics and enables the investigation of its system-size dependence. 
By comparing its predictions with numerically-exact tensor-network simulations, we identify signatures of fluctuation-induced bistability already in small systems on finite timescales. Our results establish fluctuation-induced bistability as a universal feature of dissipative cavity-coupled many-body systems and provide a general framework for the investigation of its non-equilibrium dynamics across a wide range of hybrid quantum platforms.
\end{abstract}

\author{Luisa Tolle}
\thanks{These authors contributed equally to this work.}
\affiliation{\cea}
\affiliation{\bonnpi}
\author{Simon B.~J\"{a}ger}
\thanks{These authors contributed equally to this work.}
\affiliation{\bonnpi}
\author{Ameneh Sheikhan}
\affiliation{\bonnpi}
\author{Corinna Kollath}
\affiliation{\bonnpi}
\author{Catalin-Mihai Halati}
\affiliation{\pks}
\maketitle

Strong coupling between interacting many-body systems and cavity light is a rapidly advancing frontier \cite{FriskKockumNori2019, FornDiazSolano2019, delaTorreSentef2021, MivehvarRitsch2021,  BlaisWallraff2021, SchlawinSentef2022, BretscherSentef2026}, promising novel quantum materials such as light-induced superconductors~\cite{Koppens2011,Cavalleri2018,Torre2021,KerenBasov2026, MontanaroFausti2026, ZhangGao2026}, as well as new routes toward light-driven chemistry \cite{Lidzey1998,Lidzey2000,Hutchison2011,Herrera2016,Hirai:2023} and quantum technologies with enhanced matter-matter and light-matter entanglement \cite{Leroux:2010,Pezze:2018,Xia2023,Brady2023,Luo:2024}. 
In the field of ultracold atoms, experimental realizations of strongly coupled quantum light and matter, range from bosonic atoms \cite{BaumannEsslinger2010, KlinderHemmerich2015, KlinderHemmerich2015b, LandigEsslinger2016, HrubyEsslinger2018, KroezeLev2018, VaidyaLev2018, KesslerHemmerich2021, FerriEsslinger2021, DreonDonner2022, KongkhambutKessler2022,Young2024}, to fermionic atoms \cite{RouxBrantut2020, ZhangWu2021, HelsonBrantut2023,  WuWu2023, ZwettlerBrantut2025, ZwettlerBrantut2025b, BuhlerBrantut2026} and spin degrees of freedom \cite{Masson:2014,Masson:2017,Kroeze:2018,LandiniEsslinger2018,SauerweinBrantut2023}.
The theoretical description of these systems is highly challenging, as it requires capturing the dynamics of many interacting particles coupled to a photonic mode \cite{MivehvarRitsch2021}. Even state-of-the-art approaches, such as Keldysh path integrals \cite{GopalakrishnanGoldbart2009, PiazzaStrack2014} or matrix product states \cite{WallRey2016, HalatiKollath2020, HalatiKollath2020b, ChiriacoChanda2022}, remain limited to relatively small system sizes. 
Consequently, many works rely on mean-field treatments, often justified by the effectively global nature of photon-mediated interactions \cite{RitschEsslinger2013, MivehvarRitsch2021}. While this approximation can become exact for Dicke-like systems~\cite{KirtonDallaTorre2019, CarolloLesanovsky2021}, its validity for interacting systems is far less clear. 
Although mean-field theory provides valuable insight into emergent phases and short-time dynamics, the role of genuine beyond-mean-field effects remains a fundamental open question \cite{MivehvarRitsch2021, GopalakrishnanGoldbart2009, PiazzaStrack2014, DamanetKeeling2019, HalatiKollath2020, HalatiKollath2020b, BezvershenkoRosch2021, HalatiKollath2022, TolleHalati2025, TolleHalati2026, ChiriacoChanda2022, JagerBetzholz2022, LinkDaley2022, HalatiKollath2025, MuellerStrunz2025, HalatiJager2025, OrsoDeuar2025, SchmitJaeger2026}.

\begin{figure}[!hbtp]
	\centering
	\includegraphics[width=0.48\textwidth]{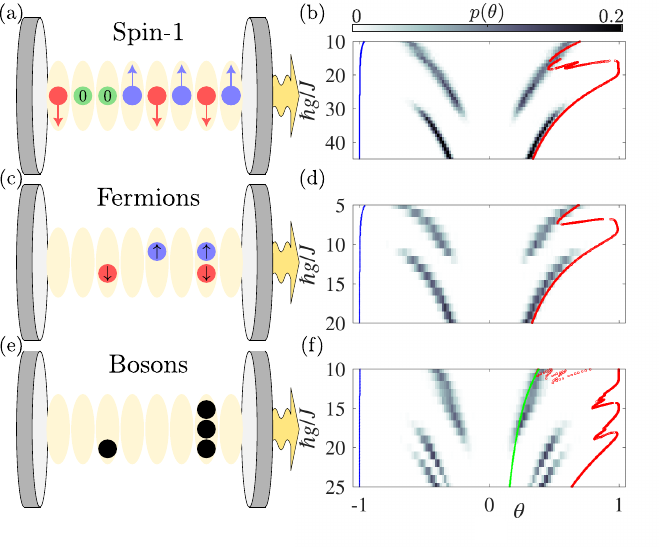}
	\caption{Sketch of (a) the spin-1 system, (c) spinful fermions, and (e) spinless bosons interacting with a cavity. Distributions $p(\theta)$ obtained from DSRE are displayed in panels (b), (d), and (f) for spin-1, fermions, and bosons, respectively.  We used the parameters (b) $U_0 = 30J$, $\hbar\delta = 10J$, $\hbar \Gamma = 3J$, $L = 224$, total magnetization $\sum_j\langle\hat{S}^z_j\rangle=-L/2$; (d) $U_0 = 20J$, $\hbar\delta = 5J$, $\hbar\Gamma = 3J$, $L = 224$, with quarter filling $N_\downarrow = L/4$, $N_\uparrow = L/4$; and (f) $U_0 = 40J$, $\hbar\delta = 5J$, $\hbar \Gamma = 3J$, $L = 224$, with half filling $N =L/2$. The red lines in (b), (d), (f) show $|\theta_{\mathrm{th}}|$ obtained for $L=8$ from the MF+fluct. method. The green data points are obtained for onsite boson number cutoff $N_{\mathrm{max}}=1$. The blue lines are $-|\theta_{\mathrm{MF}}|$.}
	\label{fig:1}
\end{figure}

Recently, a framework incorporating quantum fluctuations beyond mean field has enabled access to non-equilibrium steady states beyond conventional approaches~\cite{BezvershenkoRosch2021,HalatiKollath2022,TolleHalati2025,TolleHalati2026}. A striking prediction is fluctuation-induced bistability, which emerges from the interplay between dissipation-driven fluctuations and many-body energy scales and is therefore absent at the mean-field level~\cite{TolleHalati2025,TolleHalati2026}. This mechanism is fundamentally distinct from the bistability associated with the coexistence of different phases in systems with photon-mediated long-range interactions~\cite{LandigEsslinger2016,Flottat:2017,Himbert:2019,Wu:2024}. Whether fluctuation-induced bistability is a genuine many-body phenomenon or merely an artifact of the underlying approximation has remained an open question. Resolving this issue is essential for establishing it as a generic consequence of strong light-matter and matter-matter interactions and for identifying experimentally accessible regimes in which it can be observed.

In this work, we show that fluctuation-induced bistability, originally identified for fermionic particles coupled to a dissipative photonic mode \cite{TolleHalati2025}, arises in a broad class of hybrid many-body systems coupled to cavities, including bosonic and spin systems (see Fig.~\ref{fig:1}). 
The key ingredient for the fluctuation-induced bistability is the presence of strong inter-particle interactions, whose interplay with the coupling to the cavity leads to efficient energy transfer channels, that can stabilize macroscopically distinct states. 
These bistabilities are present only if quantum fluctuations in the coupling between matter and photons are taken into account. 
To connect to experimentally observable quantities, we go beyond the discussion of steady-state results and analyze the signatures of the fluctuation-induced bistability in the non-equilibrium dissipative dynamics of the hybrid system. To achieve this, we developed an approach relying on a polaron transformation of the matter-cavity coupling. We show that the bistable behavior can already be identified at intermediate times and dominate the dynamics.

\textit{Models and Lindblad equation}-- We consider systems  which consist of a dissipative cavity mode globally coupled to an interacting many-body system (see Fig.~\ref{fig:1}). 
These systems can be described by a many-body Lindblad master equation \cite{GoriniSudarshan1976, Lindblad1976, CarmichaelBook, BreuerPetruccione2002, MaschlerRitsch2008, RitschEsslinger2013, MivehvarRitsch2021}
\begin{align}
	\label{eq:Lindblad}
	\frac{\partial \hat{\rho}}{\partial t} = -\frac{i}{\hbar}\left[ \hat{H}, \hat{\rho} \right] + \frac{\Gamma}{2}\mathcal{D}[\hat{a}] \hat{\rho},
\end{align}
with $\mathcal{D}[\hat{J}]\hat{\rho} = 2\hat{J}\hat{\rho}\hat{J}^\dagger - \hat{J}^\dagger \hat{J}\hat{\rho} - \hat{\rho}\hat{J}^\dagger \hat{J}$, 
photon loss rate $\Gamma$, and photonic annihilation (creation) operators $\hat{a}$ ($\hat{a}^\dagger$).
The Hamiltonian reads $\hat{H}=\hat{H}_{\mathrm{mb}}+\hat{H}_{\mathrm{p}}+\hat{H}_{\mathrm{mb-p}}$, with $\hat{H}_{\mathrm{p}}=\hbar\delta \hat{a}^\dagger \hat{a}$ (detuning $\delta$), $\hat{H}_{\mathrm{mb-p}}=-\hbar g(\hat{a}+\hat{a}^\dagger)\hat{\Theta}/\sqrt{L^d}$ (light-matter coupling strength $g$, system size $L^d$, dimension $d$, matter operator $\Theta$), and many-body Hamiltonian $\hat{H}_{\mathrm{mb}}$. 
We decompose $\hat{H}_{\mathrm{mb}}=\hat{H}_{\parallel}+\hat{H}_{\perp}$, with $\hat{H}_{\parallel}$ containing all terms commuting with $\hat{\Theta}$ and $\hat{H}_{\perp}=\hat{H}_{\mathrm{mb}}-\hat{H}_{\parallel}$.

To demonstrate the generality of our approach, we consider three models (Fig.~\ref{fig:1}): the Bose-Hubbard model with bosonic operators $\hat{b}_j^{(\dagger)}$ and densities $\hat{n}_j=\hat{b}_j^\dagger\hat{b}_j$, the Fermi-Hubbard model with fermionic operators $\hat{c}_{j,\sigma}^{(\dagger)}$ and densities $\hat{n}_{j,\sigma}=\hat{c}_{j,\sigma}^\dagger\hat{c}_{j,\sigma}$, and the XX spin-1 model with spin operators $\hat{S}_j^z,\hat{S}_j^\pm$ and on-site $\hat{S}^z_j\hat{S}^z_j$ interactions. 
The models are summarized in the following table, with the matter operator $\Theta$ corresponding to odd-even density imbalance: 
\begin{table}[h!]
\flushleft
	\begin{tabular}{|c| c| c| c|}
		\hline
		&$\hat{H}_{\parallel}/U$&$\hat{H}_{\perp}/J$& $\hat{\Theta}$ \\
		\hline\hline
		Spins&$\sum_j\hat{S}_j^z\hat{S}_j^z $ &$-\sum_{\langle j,j'\rangle}\hat{S}_j^+\hat{S}_{j'}^{-}$ &$\sum_j(-1)^j\hat{S}_j^z$\\
		Fermions&$\sum_{j} \hat{n}_{j,\uparrow}\hat{n}_{j,\downarrow}$   &$ -\sum_{\sigma,\langle j,j'\rangle} \hat{c}_{j,\sigma}^\dagger \hat{c}_{j',\sigma} $ &$ \sum_{j,\sigma}(-1)^j\hat{n}_{j,\sigma}$\\
		Bosons&$ \sum_{j} \frac{\hat{n}_{j}(\hat{n}_{j}-1)}{2}$    & $- \sum_{\langle j,j'\rangle} \hat{b}_{j}^\dagger \hat{b}_{j'}$       &$ \sum_j(-1)^j\hat{n}_j$\\
		\hline
	\end{tabular}
\end{table}

\textit{Mean field with fluctuations (MF+fluct.)}-- First we recap the mean field with fluctuations (MF+fluct.) method for matter-cavity coupled systems \cite{BezvershenkoRosch2021, HalatiKollath2022, TolleHalati2025,TolleHalati2026}, which led to the identification of the fluctuation-induced bistability for fermions in Ref.~\cite{TolleHalati2025}. The MF+fluct.~method relies on the adiabatic elimination framework \cite{Garcia-RipollCirac2009, ReiterSorensen2012, Kessler2012, PolettiKollath2013, ZanardiCamposVenuti2014, SciollaKollath2015, LangeRosch2018, LenarcicRosch2018} and the starting point is a mean-field decoupling of $H_{\mathrm{mb-p}}$. 
This yields the effective Hamiltonian for the many-body system \cite{BezvershenkoRosch2021, TolleHalati2025, TolleHalati2026}
\begin{equation}
	\label{eq:cav_MF}
	\hat{H}_\text{eff}^\text{MF}=\hat{H}_{\text{mb}}-\hbar g\lambda\hat{\Theta},
\end{equation}
with $\lambda\!=\!\langle\hat{a}\!+\!\hat{a}^\dagger\rangle/\sqrt{L^d}\!=\!2g\delta\langle\hat{\Theta}\rangle/[L^d(\delta^2\!+\!(\Gamma/2)^2)]$. Fluctuations are reintroduced perturbatively via $\delta\hat{H}_\text{c}\!=\!-\hbar g\big[(\hat{a}\!+\!\hat{a}^\dagger)/\sqrt{L^d}\!-\!\lambda\big][\hat{\Theta}-\langle\hat{\Theta}\rangle]$ 
and the particles are assumed to thermalize quickly \cite{DeMarcoKollath2022}, such that a thermal density matrix can be used, i.e.~$\hat{\rho}_\text{at}\sim e^{-\beta\hat{H}_\text{eff}^\text{MF}}$.
The steady-state temperature can be obtained from the condition of constant energy transfer~\cite{BezvershenkoRosch2021, TolleHalati2025, TolleHalati2026}, i.e.
\begin{equation}
	\label{eq:energy_transfer}
	0= \frac{\partial}{\partial t}\langle\hat{H}_{\text{eff}}^\text{MF}\rangle\propto\sum_{n,m}  \frac{|\Theta_{nm}|^2e^{\!-\beta E_{m}^\text{MF}} (E_n^\text{MF}\!-\!E_m^\text{MF})\Gamma}{\left(E_n^\text{MF}\!-\!E_m^\text{MF}\!+\!\hbar\delta\right)^2\!+\!\left(\hbar\Gamma/2\right)^2},
\end{equation}
with $\Theta_{nm}=\langle n^\text{MF}|\hat{\Theta}|m^\text{MF}\rangle$, where $|m^\text{MF}\rangle$ are the eigenstates and  $E_n^\text{MF}$ the eigenenergies of $\hat{H}_{\mathrm{eff}}^\text{MF}$. The system is characterized by self-consistently solving Eqs.~\eqref{eq:cav_MF} and \eqref{eq:energy_transfer} for $\lambda$ and $\beta$, yielding $\Theta_{\mathrm{th}}=\langle\hat{\Theta}\rangle$. This is in contrast to the zero-temperature mean-field result $\Theta_{\mathrm{MF}}$, obtained by setting $1/\beta=0$ and neglecting fluctuations.

The MF+fluct.~results are shown in red in Figs.~\ref{fig:1}(b), (d), and (f) for $d=1$, where $|\theta_{\mathrm{th}}|=|\Theta_{\mathrm{th}}|/\Theta_{\mathrm{max}}$ is displayed ($\Theta_{\mathrm{max}}=L/2$ is the maximum eigenvalue of $\hat{\Theta}$ for the given filling or total magnetization). For all three models -- spin-1, fermions, and bosons -- we observe fluctuation-induced bistability, \textit{i.e.}~the coexistence of multiple stable solutions for the same $g$ values. This demonstrates that the phenomenon, previously reported for fermions~\cite{TolleHalati2025,TolleHalati2026}, is generic and independent of particle statistics. In contrast, for all models the zero-temperature mean-field prediction, $-|\theta_{\mathrm{MF}}|=-|\Theta_{\mathrm{MF}}|/\Theta_{\mathrm{max}}$ (blue solid lines in Fig.~\ref{fig:1}), exhibits no bistability but almost maximal imbalance for the values of $g$ shown, highlighting the crucial role of fluctuations. 

The fluctuation-induced bistability emerges at sufficiently large interaction strength, $U/J$, and light-matter coupling, $\hbar g/J$, in the vicinity of the photon-assisted resonances of Eq.~\eqref{eq:energy_transfer}~\cite{TolleHalati2025,TolleHalati2026}. In small-$J$ limit, these reduce to $2\hbar g\lambda \approx pU \pm \hbar\delta$, with $p=1$ for fermions and spins, and integer $p\geq1$ for bosons.  Physically, since $\hat{H}_{\|}$ and the matter-cavity coupling are diagonal in the same basis, photon-assisted processes near resonance enable efficient energy exchange between the cavity field and the interacting matter system. The resulting cooling stabilizes multiple distinct self-consistent steady states, giving rise to the fluctuation-induced bistability.

While the general features of fluctuation-induced bistability are independent of particle statistics, the quantitative characteristics of the bistability depend on the structure of the local Hilbert space. Fermions and spin-$1$ systems [Fig.~\ref{fig:1}(b),(d)] show a single bistable region, whereas bosons exhibit a hierarchy associated with higher local occupancies [Fig.~\ref{fig:1}(f)]. For spin-$1$, weak additional features at $\hbar g/J\lesssim20$ are attributed to finite-size effects ($L=8$)~\cite{TolleHalati2025,TolleHalati2026}, whereas the multiple bistable regimes observed in the bosonic case are intrinsic and persist in the thermodynamic limit. This is confirmed by changing the on-site boson number cutoff to $N_{\mathrm{max}}=1$, where bistable branches are absent (see SM~\cite{SM} for additional information).

Overall, fluctuation-induced bistability is robust across three qualitatively different many-body systems. Its detailed structure depends on the local Hilbert space, but it is linked to resonances between photon-assisted transitions and many-body energy levels in the strong-coupling regime $U/J,\,\hbar g/J\gg1$.

\textit{Dressed-State Rate Equation (DSRE)}--
Motivated by this observation, we develop the complementary Dressed-State Rate Equation (DSRE) approach, which treats the light-matter coupling exactly via a polaron transformation \cite{BreuerPetruccione2002, Mahan2013,Marsh:2021} and $\hat{H}_{\perp}$ perturbatively. We outline the main steps while details are given in the Supplemental Material~\cite{SM}. 
Using the polaron transformation $\hat{D} = \exp\left(\left[\chi \hat{a}^\dag - \chi^* \hat{a}\right]\hat{\Theta}\right)$ with $\chi = g/[\sqrt{L^d}(\delta - i\Gamma/2)]$,
we obtain for the transformed density matrix $\tilde{\rho} = \hat{D}^\dag \hat{\rho} \hat{D}$ the Lindblad equation
\begin{align}
	\frac{\partial \tilde{\rho}}{\partial t} &= \mathcal{L}_0 \tilde{\rho} + \mathcal{L}_1 \tilde{\rho}, \quad \textrm{with}\\
	\mathcal{L}_0 \tilde{\rho} &= -\frac{i}{\hbar}\left[\hat{H}_\mathrm{p} + \hat{H}_{\|} - \hbar\delta |\chi|^2 \hat{\Theta}^2, \tilde{\rho}\right] \\
	&\quad - \frac{\Gamma}{2}\mathcal{D}[\hat{a}] \tilde{\rho} - \frac{\Gamma |\chi|^2}{2}\mathcal{D}[\hat{\Theta}] \tilde{\rho} - \Gamma \left[\hat{\Theta}, \left(\chi^* \hat{a} \tilde{\rho} - \chi \tilde{\rho} \hat{a}^\dag \right)\right],\nonumber
\end{align}
and $\mathcal{L}_1 \tilde{\rho} = -\frac{i}{\hbar}\left[\hat{D}^\dag \hat{H}_{\perp} \hat{D}, \tilde{\rho}\right]$.

In the absence of $\hat{H}_{\perp}$, any state of the form $\tilde{\rho}_{\bf n} = \ket{\Psi_{\bf n}}\bra{\Psi_{\bf n}}$ is a stationary state of the dynamics, with $\ket{\Psi_{\bf n}} = \ket{{\bf n}, \mathrm{vac}}$, where $\ket{{\bf n}}$ is an eigenstate of $\hat{\Theta}$ ($\hat{\Theta}\ket{\bf n}=\Theta_{\bf n}\ket{{\bf n}}$) and $\hat{H}_{\|}$, and $\ket{\mathrm{vac}}$ is the cavity vacuum in the displaced frame (corresponding to a coherent state in the original frame). This follows from a strong symmetry generated by $\hat{\Theta}$~\cite{AlbertJiang2014, BucaProsen2012, HalatiKollath2025}. For weak $\hat{H}_{\perp}$, this symmetry is weakly broken, leading to slow dynamics within the corresponding manifold~\cite{HalatiKollath2020b,HalatiKollath2022,HalatiKollath2025}, analogous to effective descriptions in decoherence-free subspaces~\cite{Garcia-RipollCirac2009,ReiterSorensen2012,Kessler2012,PolettiKollath2013,ZanardiCamposVenuti2014,SciollaKollath2015}. We assume the density matrix to be of the form \begin{align}\label{eq:rho_rate}
	\hat{\rho}=\hat{D}\tilde{\rho}\hat{D}^\dag= \sum_{\bf n} P_{\bf n} \ket{{\bf n}, \chi \Theta_{\bf n}}\bra{{\bf n}, \chi \Theta_{\bf n}},
\end{align} in the original frame, where $P_{\bf n}$ is the probability to find the state $\ket{{\bf n}}$ and we neglected coherences between degenerate eigenstates. The dynamics within the slow subspace is described by the rate equation
\begin{align}\label{eq:rate}
	\frac{dP_{\bf n}}{dt} = \sum_{{\bf n}'} \left[- R_{{\bf n}\to{\bf n}'} P_{\bf n} + R_{{\bf n}'\to{\bf n}} P_{{\bf n}'} \right],
\end{align}
 with $R_{{\bf n}\to{\bf n}'} = 2 \left|\bra{{\bf n}'} \hat{H}_{\perp} \ket{{\bf n}}\right|^2 \mathrm{Re}[I_{\bf n,n'}]/\hbar^2$ and
\begin{align}
	I_{\bf n,n'}=&\int_{0}^{\infty}dt\, e^{-i\Delta E_{\bf n',n}\frac{t}{\hbar}-|\chi|^2\Delta \Theta_{\bf n',n}^2\left[\frac{\Gamma t}{2}+\frac{\chi}{\chi^*}\left(1-e^{-[i\delta+\frac{\Gamma}{2}]t}\right)\right]}.
\end{align}
We introduced $\Delta E_{\bf n',n}=E_{\bf n'}-E_{\bf n}$ and $\Delta\Theta_{\bf n',n}=\Theta_{\bf n'}-\Theta_{\bf n}$, with $E_{\bf n}\ket{{\bf n}}=(\hat{H}_{\|}-\hbar\delta |\chi|^2\hat{\Theta}^2)\ket{\bf n}$. 
In the following, we perform dynamical Monte Carlo simulations of Eq.~\eqref{eq:rate}, which enable access to large system sizes.

In Figs.~\ref{fig:1}(b),(d),(f), we show the probabilities $p(\theta)$ to find the system at a given $\theta=\Theta_{\bf n}/\Theta_{\mathrm{max}}$,
\begin{align}
	p(\theta) = \sum_{{\bf n}} P_{\bf n}\delta_{\theta,\Theta_{\bf n}/\Theta_{\mathrm{\max}}},
\end{align}
with Kronecker-delta $\delta_{\theta,\Theta_{\bf n}/\Theta_{\mathrm{\max}}}$ and
obtained for spin-1, fermions, and bosons after a finite time $Jt/\hbar=1000$. 
Initial states are sampled: for fermions and bosons by distributing particles over the lattice with at most one per site, for spins the initial state is prepared in eigenstates of $\hat{S}_j^z$ of which $L/2$ have eigenvalues $0$ and $L/2$ have eigenvalue $-1$. In all cases, $p(\theta)$ is initially centered at $\theta=0$, from which the dynamics starts.
For all three systems, DSRE simulations show the fluctuation-induced bistability after finite time. For spin-$1$ and fermions, a single bistable region appears around $\hbar g/J\approx30$ and $\hbar g/J\approx12$, respectively, while bosons exhibit two regions around $\hbar g/J\approx17$ and $\hbar g/J\approx22$. Overall, DSRE and MF+fluct.~agree qualitatively, though MF+fluct.~predicts narrower bistable regions of smaller $\hbar g/J$. The largest deviations occur for bosons, where DSRE yields broader bistability, while MF+fluct.~without a local atom number cutoff has larger $\theta_{\mathrm{th}}$. The agreement is better with a cutoff $N_{\mathrm{max}}=1$ (green) for MF+fluct., suggesting that DSRE at finite time favors low-density-imbalance states with lower local occupancies. 

\begin{figure}[!hbtp]
	\centering
	\includegraphics[width=0.48\textwidth]{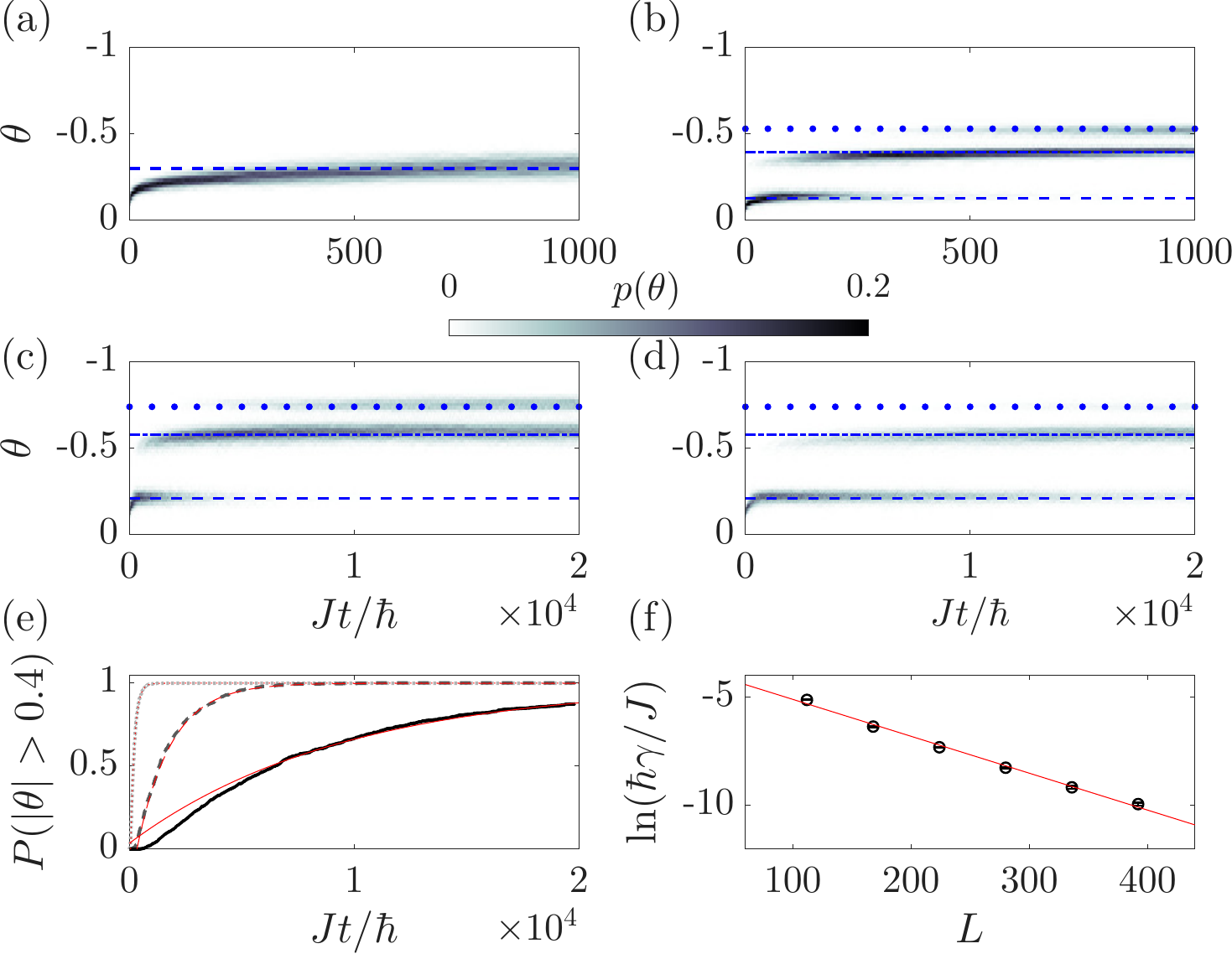}
	\caption{(a)-(d) Time evolution of $p(\theta)$ (bosons) for four parameter sets: (a) $L = 224$, $\hbar g = 14J$, (b) $L = 224$, $\hbar g = 22J$, (c) $L = 224$, $\hbar g = 18J$, and
	(d) $L = 336$, $\hbar g = 18J$.
	Guide to the eye: blue dashed lines in (a),(b),(c),(d) indicate positions of first peak, dashed-dotted line in (b),(c),(d) second peak and the dotted line in (b),(c),(d) third peak.	(e) Probability $P(|\theta| > 0.4)$
	as function of time for atom numbers $L = 112$ (light gray), $L = 224$ (dark gray), and $L = 336$ (black) at $\hbar g = 18J$. The red curves are fits $1-\exp(-\gamma t)$. (f) Growth rate $\gamma$, extracted from the fits in panel (e), plotted as a function of $N$. The red line is a linear fit to $\log(\hbar\gamma/J)$. Remaining parameters are $U_0 = 40J$, $\hbar\delta = 5J$, $\hbar\Gamma = 3J$, and $N = L/2$.
	}
	\label{fig:2}
\end{figure}

The described differences originate from the distinct assumptions: unlike MF+fluct., DSRE does not assume thermalization, but treats tunneling perturbatively and resolves real-time dynamics rather than steady states.
The light-matter correlations present in the DSRE state from Eq.~\eqref{eq:rho_rate} are 
consistent with semiclassical~\cite{SchuetzMorigi2013}, quantum~\cite{HalatiKollath2020,HalatiKollath2020b}, and experimental results~\cite{BuhlerBrantut2026}. In particular, Eq.~\eqref{eq:rho_rate} allows access to the full statistics of $\theta$ via the cavity field, e.g., through the Glauber-Sudarshan $P$ representation~\cite{GardinerZollerBook}, $P(\chi\Theta_{\mathrm{max}}\theta)=p(\theta)$.

\textit{Dynamics}-- A central achievement of the DSRE method is the simulation of real time dynamics that we show in the following for the bosonic model and the previously discussed initial state at half filling. For the fermionic and spin models we refer to the SM~\cite{SM}. 
Figure~\ref{fig:2} shows the dynamics of $p(\theta)$ for $Jt/\hbar\leq1000$, $L=224$, and (a) $\hbar g=14J$ (no bistability), (b) $\hbar g=22J$ (bistability). In both cases, the distribution rapidly evolves from $\theta=0$ to peaks at finite $\pm|\theta|$ (only negative peaks are shown due to symmetry). This is followed by slower evolution where the peaks converge to larger $|\theta|$. In (a), outside the bistable regime, the distribution relaxes to a single stationary value of $|\theta|$ (blue dashed line). In contrast, Fig.~\ref{fig:2}(b), which corresponds to the bistable regime of Fig.~\ref{fig:1}(f), shows a multistep evolution: after quick relaxation ($|\theta|\approx0.13$, blue dashed line), additional peaks emerge at $|\theta|\approx0.39$ (dash-dotted) and finally $|\theta|\approx0.53$ (dotted). These results emphasize the bistable behavior, and reveal strongly parameter-dependent dynamics with qualitatively different pathways towards steady state.

\begin{figure}[!hbtp]
	\centering
	\includegraphics[width=0.48\textwidth]{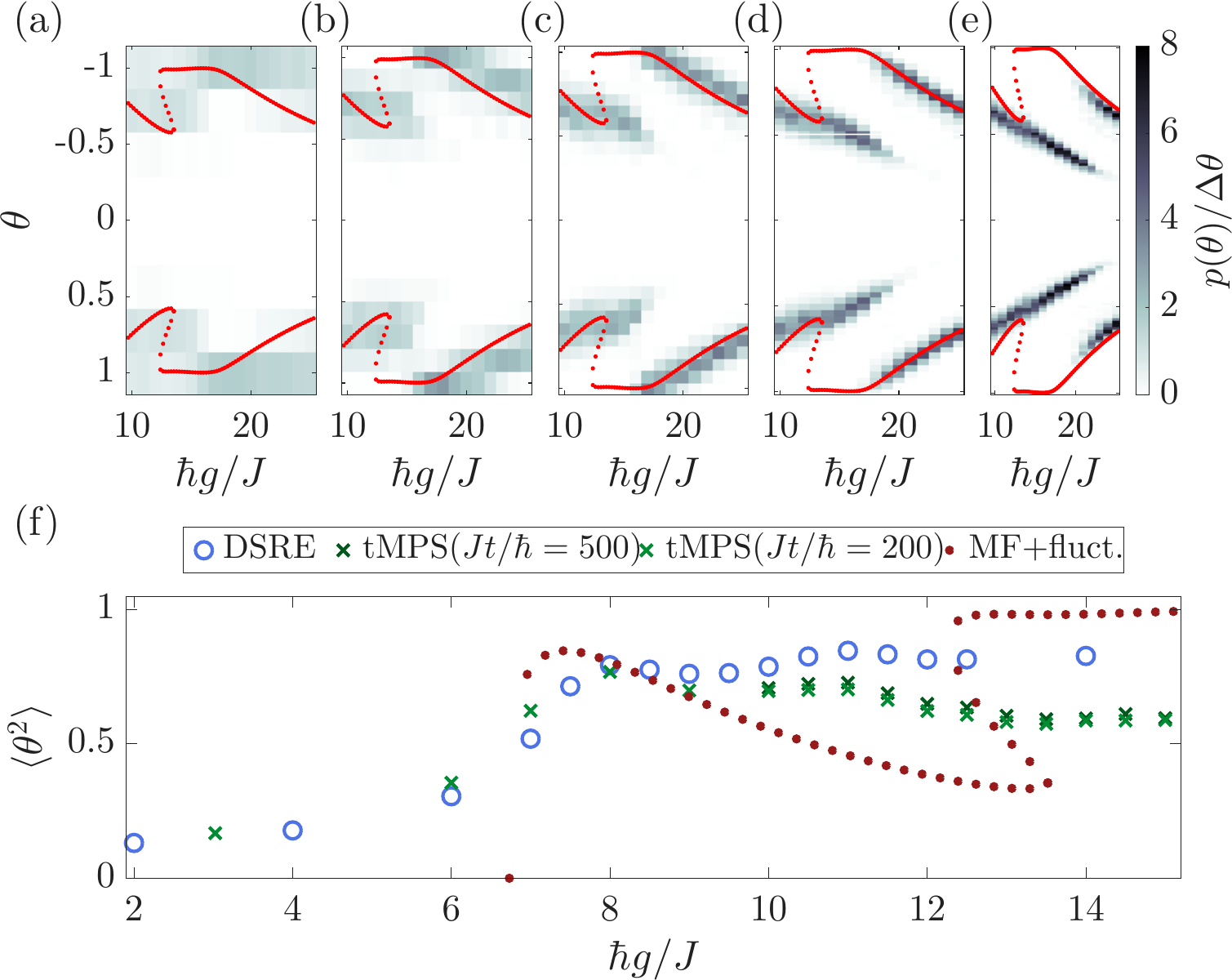}
	\caption{(a)-(e) Probability density $p(\theta)$ after a time $Jt/\hbar=2\times10^4$ as function $g$ for (a) $L=14$, (b) $L=28$ (c) $L=56$ (d) $L=112$ (e) $L=224$ for bosons with $N_{\mathrm{max}}=2$. Red dots show the solution of MF+fluct.. (f) Expectation value $\langle\theta^2\rangle$ extracted from the steady state of Eq.~\eqref{eq:rate} (blue circles), tMPS simulations for $L=14$ (dark green crosses $Jt/\hbar=500$, light green crosses $Jt/\hbar=200$), and MF+fluct.~in red. Remaining parameters are $U_0 = 40J$, $\hbar\delta = 10J$, $\hbar\Gamma = 3J$, $N = L/2$.
	}
	\label{fig:3}
\end{figure}

With DSRE, we observe a strong system-size dependence of the slow dynamics. Figures~\ref{fig:2}(c),(d) show results for $L=224$ and $L=336$, respectively. For larger systems, the emergence of high-$|\theta|$ peaks is delayed, while low-$|\theta|$ peaks persist much longer before eventually diminishing. In contrast, the smaller system exhibits shorter-lived low-$|\theta|$ peaks and an additional third peak at larger $|\theta|$. Despite these differences in dynamics, the positions of the low- and high-$|\theta|$ peaks remain largely unchanged [blue lines in (c),(d)]. 

To quantify the peak dynamics, we compute the weight of higher peaks by $P(|\theta|>0.4)=\sum_{|\theta|>0.4}p(\theta)$ for different parameters in Fig.~\ref{fig:2}(e). In all cases, $P(|\theta|>0.4)\to1$ at long times, showing that the low-$|\theta|$ peak is only metastable at finite $L$, while relaxation slows down strongly with increasing system size.
To characterize this, we fit $P(|\theta|>0.4)\sim 1-\exp(-\gamma t)$ and extract the rate $\gamma$ [Fig.~\ref{fig:2}(f)]. The data follow an approximately linear trend on a logarithmic scale, indicating an exponential decay $\gamma\propto a^{-L}$ and, thus, exponentially growing relaxation times \cite{Znidaric2015,DefenuTrombettoni2023}. This strong slowdown reflects that larger systems require more intermediate processes to reach high-$|\theta|$ states, making the low-$|\theta|$ peak increasingly long-lived. In the thermodynamic limit, this suggests true stabilization of the low-$|\theta|$ peak and an enlarged bistable regime.

\textit{Scaling with system size}-- To show how the distribution changes as function of $L$ we perform DSRE simulations for different values of $L$, fixing $N=L/2$ and time $Jt/\hbar=2\times10^4$ visible in Fig.~\ref{fig:3}. 
A local cutoff of $N_{\mathrm{max}}=2$ is used for simplicity. 
For small systems ($L\le 28$), the bistable region is barely visible due to strong fluctuations and the coarse discretization $\Delta\theta=4/L$. For intermediate sizes ($56\le L\le112$), a clear two-peak structure emerges over a broad range of couplings signaling the bistable region. For $L=56,112$ [Figs.~\ref{fig:3}(c),(d)], we find reasonable agreement in the overall structure with steady-state MF+fluct.~calculations with the same cutoff $N_{\mathrm{max}}=2$. For larger systems ($L=224$), and within accessible timescales, the low-$|\theta|$ peak persists to higher couplings while the high-$|\theta|$ peak loses weight at lower couplings. This reflects the slow probability transfer between peaks, consistent with Fig.~\ref{fig:2}(e).

\textit{Fluctuation-induced bistability in the exact quantum dynamics}--
In Fig.~\ref{fig:3}(a)-(e), we find that true bistability emerges only for sufficiently large systems, $L>28$. Nevertheless, precursor signatures of bistability, such as the non-monotonic behavior of $\theta$ [see Fig.~\ref{fig:3}(a)], are already visible for $L=14$. This observation enables us to explore these signatures employing numerically exact real-time evolution of Eq.~\eqref{eq:Lindblad} using a recently developed method based on time-dependent matrix product state (tMPS) simulations~\cite{HalatiKollath2020,HalatiKollath2020b,TolleHalati2026b,SM}. Figure~\ref{fig:3}(f) compares the resulting values of $\langle\theta^2\rangle$ obtained from the different approaches~\cite{SM}.
For small coupling $g$, we observe very good agreement between tMPS and DSRE results, both exhibiting a similar onset near $\hbar g \approx 6J$, consistent with the MF+fluct.~prediction. 
Beyond this point, we obtain a qualitative agreement between the finite time tMPS results and the steady state DSRE.
In particular, the tMPS results for the imbalance show the same non-monotonic dependence on $g$ as obtained for DSRE, hinting at the emergence of the bistable region for larger system sizes.  
This is further supported by the slow evolution obtained in tMPS around the bistable region, consistent with the DSRE results~\cite{SM}.

For both the tMPS and DSRE results, the bistability does not manifest as two distinct solutions at finite $L$, but through a density matrix containing a mixture of states with different imbalance [as discussed around Eq.~\eqref{eq:rho_rate}]. This explains why the value of $\langle\theta^2\rangle$ for both approaches lies between the two solutions obtained with MF+fluct..
Furthermore, we can use the comparison between the exact numerics and DSRE to understand the processes which can become relevant beyond Eqs.~\eqref{eq:rho_rate}-\eqref{eq:rate}.
While DSRE assumes infinitesimal tunneling, tMPS captures finite tunneling effects which compete with the coupling to the cavity $\propto \hat{\Theta}$, thereby reducing $\langle \theta^2 \rangle$. 
Furthermore, the density matrix Ansatz used in Eq.~\eqref{eq:rho_rate}, does not contain coherences between states  ${\bf n}\neq{\bf n}'$ with same $\Theta_{\bf n}=\Theta_{\bf n'}$ but which are long-lived in the presence of the approximate strong symmetry \cite{HalatiKollath2025}.

\textit{Conclusion}-- We have demonstrated that fluctuation-induced bistability is a robust and generic phenomenon in interacting many-body systems coupled to a dissipative cavity by developing and applying the MF+fluct.~and DSRE approaches to spins, fermions, and bosons. This bistability emerges from the interplay of interactions, dissipation, and fluctuations, and is linked to resonances between photonic transitions and many-body energy scales, a mechanism common to all the microscopic models considered. 
Furthermore, the DSRE method enables the study of large systems and long timescales, revealing rich size-dependent dynamics. Numerically-exact tMPS simulations show that the signatures of fluctuation-induced bistability are already observable in small systems, demonstrating that the phenomenon emerges dynamically and persists across system sizes. 
Our work opens several promising research directions based on the developed methods, including the study of dynamics and steady states in different spatial dimensions and various physical setups.
More broadly, we believe that our work provides an important stepping stone toward understanding non-equilibrium dynamics arising from the interplay of dissipation, fluctuations, many-body interactions, and light-matter coupling.

\emph{Acknowledgments:}
We are deeply grateful to Thierry Giamarchi for insightful discussions and contributions, in particular to the MF+fluct.~based part of this work, and for the careful reading of the manuscript. We further thank Helmut Ritsch, Farokh Mivehvar, Giovanna Morigi for fruitful discussions.
We acknowledge support by the Deutsche Forschungsgemeinschaft (DFG, German Research Foundation) under Project No.~277625399-TRR 185 OSCAR (``Open System Control of Atomic and Photonic Matter'', B3 and B4), No.~277146847-CRC 1238 (``Control and dynamics of quantum materials'', C05), CRC 1639 NuMeriQS (``Numerical Methods for Dynamics and Structure Formation in Quantum Systems'') – project
No.~511713970,
and under Germany’s Excellence Strategy – Cluster of Excellence Matter and Light for Quantum Computing (ML4Q) EXC 2004/1 – 390534769.

\emph{Data availability:} The supporting data for this article are openly available at Zenodo \cite{datazenodo}.

\pagebreak

\begin{widetext}

\section{Supplemental Material}

\setcounter{section}{0}
\renewcommand{\thesection}{\Alph{section}}
\setcounter{section}{0}
\renewcommand{\thesubsection}{\arabic{subsection}}

\setcounter{equation}{0}
\renewcommand{\theequation}{S.\arabic{equation}}
\setcounter{figure}{0}
\renewcommand{\thefigure}{S\arabic{figure}}

\section{Derivation of the dressed state rate equation}
In this section we show the detailed derivation of the dressed state rate equation (DSRE).
Our starting point is the master equation for $\tilde{\rho}=\hat{D}^\dag\hat{\rho}\hat{D}$ with $\hat{D}=\exp\left(\left[\chi\hat{a}^\dag-\chi^*\hat{a}\right]\hat{\Theta}\right)$, and $\chi=g/[\sqrt{L^d}(\delta-i\Gamma/2)]$ reported in the main text. This master equation is given by
\begin{align}
	\frac{\partial\tilde{\rho}}{\partial t}=&\mathcal{L}_0\tilde{\rho}+\mathcal{L}_1\tilde{\rho}.\label{eq:master}
\end{align}
Here, $\mathcal{L}_0$ given by
\begin{align}
\mathcal{L}_0\tilde{\rho}=&-\frac{i}{\hbar}\left[\hat{H}_\mathrm{cav}+\hat{H}_{\|}-\delta|\chi|^2\hat{\Theta}^2,\tilde{\rho}\right]\\
	&-\frac{\Gamma}{2}\mathcal{D}[\hat{a}]\tilde{\rho}-\frac{\Gamma|\chi|^2}{2}\mathcal{D}[\hat{\Theta}]\hat{\rho}-\Gamma\left[\hat{\Theta},\left(\chi^*\hat{a}\tilde{\rho}-\chi\tilde{\rho}\hat{a}^\dag\right)\right]\nonumber
\end{align}
determines the fast time scale and $\mathcal{L}_1$ given by
\begin{align} \mathcal{L}_1\tilde{\rho}=-\frac{i}{\hbar}\left[\hat{D}^\dag\hat{H}_{\perp}\hat{D},\tilde{\rho}\right]
\end{align}
determines the slow timescale which is in the main paper proportional to the tunneling $J$.

\subsection{Projector method}
To eliminate the fast degrees of freedom we follow the Mori-Zwanzig formalism and introduce the projector
\begin{align}
	\mathcal{P}\tilde{\rho}=\sum_{\bf n}\ket{\Psi_{\bf n}}\bra{\Psi_{\bf n}}\tilde{\rho}\ket{\Psi_{\bf n}}\bra{\Psi_{\bf n}}=\sum_{\bf n}P_{\bf n}\hat{\rho}_{\bf n}
\end{align}
onto the described state space spanned by $\hat{\rho}_{\bf n}=\ket{\Psi_{\bf n}}\bra{\Psi_{\bf n}}$ with $\ket{\Psi_{\bf n}}=\ket{\bf n,\mathrm{vac}}$, introduced in the main text. In addition $\mathcal{Q}=1-\mathcal{P}$ is the projector on the orthogonal subspace. Now applying $\mathcal{P}$ and $\mathcal{Q}$ onto Eq.\eqref{eq:master} we derive the coupled differential equations
\begin{align}
	\frac{\partial\hat{v}}{\partial t}=&\mathcal{P}\mathcal{L}_0\hat{v}+\mathcal{P}\mathcal{L}_1\hat{w},\label{eq:dvdt}\\
		\frac{\partial\hat{w}}{\partial t}=&\mathcal{Q}\mathcal{L}_0\hat{w}+\mathcal{Q}\mathcal{L}_1\hat{w}+\mathcal{Q}\mathcal{L}_1\hat{v},
\end{align}
with $\hat{v}=\mathcal{P}\tilde{\rho}$ and $\hat{w}=\mathcal{Q}\tilde{\rho}$. 

We now describe the dynamics of $\hat{v}$ on a coarse-grained timescale $\Delta t$, which is taken to be much longer than the typical relaxation timescale associated with $\mathcal{L}_0$, but still sufficiently short to be regarded as quasi-infinitesimal on the timescale over which $\hat{v}$ evolves. We can then solve for $\hat{w}$ using
\begin{align}
	\hat{w}(t)=&e^{\mathcal{Q}[\mathcal{L}_0+\mathcal{L}_1]\Delta t}\hat{w}(t-\Delta t)+\int_{t-\Delta t}^{t}\,d\tau\,e^{\mathcal{Q}[\mathcal{L}_0+\mathcal{L}_1](t-\tau)}\mathcal{Q}\mathcal{L}_1\hat{v}(\tau).
\end{align}
The first term can be neglected since $\Delta t$ is assumed to be sufficiently long compared to the relaxation timescale associated with $\mathcal{L}_0$, such that it has decayed to zero. The second term can then be simplified as
\begin{align}
	\hat{w}(t)\approx&\int_{0}^{\infty}\,d\tau\,e^{\mathcal{Q}[\mathcal{L}_0+\mathcal{L}_1](t-\tau)}\mathcal{Q}\mathcal{L}_1\hat{v}(\tau)\\
		\approx&\int_{0}^{\infty}\,d\tau\,e^{\mathcal{Q}\mathcal{L}_0\tau}\mathcal{Q}\mathcal{L}_1\hat{v}(t)+\int_{0}^{\infty}\,d\tau\,e^{\mathcal{Q}\mathcal{L}_0\tau}\mathcal{Q}\mathcal{L}_1\int_{0}^{\infty}\,d\tau'\,e^{\mathcal{Q}\mathcal{L}_0\tau'}\mathcal{Q}\mathcal{L}_1\hat{v}(t),\label{eq:w}
\end{align}
where we have made the substitution $\tau \mapsto t-\tau$, invoked the slow evolution of $\hat{v}$ to approximate $\hat{v}(t-\tau)\approx\hat{v}(t)$, noting that the resulting corrections are of higher order in $\mathcal{L}_1$, and finally taken the limit $\Delta t\to\infty$, assuming that $e^{\mathcal{Q}\mathcal{L}_0 t}$ decays to zero on the coarse-graining timescale $\Delta t$.

Using the result of Eq.~\eqref{eq:w} in Eq.~\eqref{eq:dvdt} yields
\begin{align}
	\frac{\partial\hat{v}}{\partial t}\approx&\mathcal{P}\mathcal{L}_1\int_{0}^{\infty}d\tau e^{\mathcal{Q}\mathcal{L}_0\tau}\mathcal{Q}\mathcal{L}_1\hat{v}(t)+\mathcal{P}\mathcal{L}_0\int_{0}^{\infty}d\tau e^{\mathcal{Q}\mathcal{L}_0\tau}\mathcal{Q}\mathcal{L}_1\int_{0}^{\infty}d\tau' e^{\mathcal{Q}\mathcal{L}_0\tau'}\mathcal{Q}\mathcal{L}_1\hat{v}(t),
\end{align}
where we dropped terms that are higher than second order in $\mathcal{L}_1$.

In order to derive the DSRE, we project the effective equation of motion onto the diagonal basis,
\begin{align}
	\frac{dP_{\mathbf n}}{dt}
	=
	\mathrm{Tr}\!\left[
	\hat{\rho}_{\mathbf n}
	\frac{\partial \hat{v}}{\partial t}
	\right].
\end{align}
To this end, we evaluate separately the incoming ($I$) and outgoing ($O$) contributions,
\begin{align}
	\frac{\partial P_{\bf n}}{\partial t}=&I+O,\\
	O=&-\frac{1}{\hbar^2}\mathrm{Tr}\left[\hat{\rho}_{\bf n}\hat{D}^\dag\hat{H}_\perp\hat{D}\left\{\int_{0}^{\infty}d\tau e^{\mathcal{Q}\mathcal{L}_0\tau}\left[\sum_{\bf m}P_{\bf m}\hat{D}^\dag\hat{H}_\perp\hat{D}\ket{\Psi_{\bf m}}\bra{\Psi_{\bf m}}\right]\right\}\right]+\mathrm{c.c.},\\
	I=&\frac{1}{\hbar^2}\mathrm{Tr}\left[\hat{\rho}_{\bf n}\left\{\int_{0}^{\infty}d\tau e^{\mathcal{Q}\mathcal{L}_0\tau}\left[\sum_{\bf m}P_{\bf m}\hat{D}^\dag\hat{H}_\perp\hat{D}\ket{\Psi_{\bf m}}\bra{\Psi_{\bf m}}\right]\right\}\hat{D}^\dag\hat{H}_\perp\hat{D}\right]+\mathrm{c.c.}\\
	&+\frac{\Gamma}{\hbar^2}\mathrm{Tr}\left[\hat{\rho}_{\bf n}\hat{a}\left\{\int_{0}^{\infty}d\tau e^{\mathcal{Q}\mathcal{L}_0\tau}\left[\left\{\int_{0}^{\infty}d\tau' e^{\mathcal{Q}\mathcal{L}_0\tau'}\left[\sum_{\bf m}P_{\bf m}\hat{D}^\dag\hat{H}_\perp\hat{D}\ket{\Psi_{\bf m}}\bra{\Psi_{\bf m}}\right]\right\}\hat{D}^\dag\hat{H}_\perp\hat{D}\right]\right\}\hat{a}^\dag\right]+\mathrm{c.c.}\,.
\end{align}
For the last line in $I$ we can use
\begin{align}
	\int_{0}^{\infty}e^{\mathcal{Q}\mathcal{L}^\ddag_0\tau}\hat{a}^\dag\hat{\rho}_{\bf n}\hat{a}=\frac{1}{\Gamma}\sum_{n_{\mathrm{phot}}=1}^{\infty}\ket{{\bf n},n_{\mathrm{phot}}}\bra{{\bf n},n_{\mathrm{phot}}},
\end{align}
where $\mathcal{L}_0^\ddag$ is the adjoint of $\mathcal{L}_0$, and $n_\mathrm{phot}$ denotes the Fock states of the photons. We can then simplify 
\begin{align}
	I=&\frac{1}{\hbar^2}\sum_{n_{\mathrm{phot}=0}}^\infty\bra{{\bf n},n_{\mathrm{phot}}}\left\{\int_{0}^{\infty}d\tau e^{\mathcal{Q}\mathcal{L}_0\tau}\left[\sum_{\bf m}P_{\bf m}\hat{D}^\dag\hat{H}_\perp\hat{D}\ket{\Psi_{\bf m}}\bra{\Psi_{\bf m}}\right]\right\}\hat{D}^\dag\hat{H}_\perp\hat{D}\ket{{\bf n},n_{\mathrm{phot}}}+\mathrm{c.c}\\
	=&\frac{1}{\hbar^2}\mathrm{Tr}_{\mathrm{phot}}\left[\bra{{\bf n}}\left\{\int_{0}^{\infty}d\tau e^{\mathcal{Q}\mathcal{L}_0\tau}\left[\sum_{\bf m}P_{\bf m}\hat{D}^\dag\hat{H}_\perp\hat{D}\ket{\Psi_{\bf m}}\bra{\Psi_{\bf m}}\right]\right\}\hat{D}^\dag\hat{H}_\perp\hat{D}\ket{{\bf n}}\right]+\mathrm{c.c},
\end{align}
where  $\mathrm{Tr}_{\mathrm{phot}}$ is the trace over the photonic degrees of freedom. For $O$ we get
\begin{align}
	O=&-\frac{1}{\hbar^2}\bra{{\bf n},\mathrm{vac}}\hat{D}^\dag\hat{H}_\perp\hat{D}\left\{\int_{0}^{\infty}d\tau e^{\mathcal{Q}\mathcal{L}_0\tau}\left[\sum_{\bf m}P_{\bf m}\hat{D}^\dag\hat{H}_\perp\hat{D}\ket{\Psi_{\bf m}}\bra{\Psi_{\bf m}}\right]\right\}\ket{{\bf n},\mathrm{vac}}+\mathrm{c.c}.\,.
\end{align}

\subsection{Time evolution from $\mathcal{L}_0$}

In what follows we simplify the expressions using the explicit form of $\mathcal{L}_0$. 
Note that
\begin{align}
	\hat{D}\ket{\Psi_{\bf n}}=&	\hat{D}(\chi\Theta_{\bf n})\ket{\Psi_{\bf n}}=\ket{{\bf n},\chi\Theta_{\bf n}}
\end{align}
where the last term, $\chi\Theta_{\bf n}$, denotes the corresponding coherent state. We can use this and show
\begin{align}
	\hat{D}^\dag\hat{H}_\perp\hat{D}\ket{\Psi_{\bf n}}\bra{\Psi_{\bf n}}=\sum_{\bf n'}\ket{{\bf n'},\chi(\Theta_{\bf n}-\Theta_{\bf n'})}\bra{\bf n'}\hat{H}_\perp\ket{\bf n}\bra{\Psi_{\bf n}}
\end{align}
by introducing an identity on the particles' Hilbert space $\sum_{\bf n'}\ket{\bf n'}\bra{\bf n'}$.

To perform the time evolution we make the ansatz
\begin{align}
	\hat{\rho}(t)=C_{{\bf n}',{\bf n}}(t)\ket{{\bf n}',\alpha_{{\bf n}',{\bf n}}(t)}\bra{\Psi_{\bf n}},
\end{align}
with initial condition $C_{\bf n',n}(0)=1$ and $\alpha_{{\bf n}',{\bf n}}(0)=\chi(\Theta_{\bf n}-\Theta_{\bf n'})$. We now perform the explicit derivative 
\begin{align}
	\frac{\partial\hat{\rho}}{\partial t}=&d_tC_{{\bf n}',{\bf n}}\ket{{\bf n}',\alpha_{{\bf n}',{\bf n}}}\bra{\Psi_{\bf n}}+\left[d_t\alpha\hat{a}^\dag-\frac{d_t|\alpha|^2}{2}\right]C_{{\bf n}',{\bf n}}\ket{{\bf n}',\alpha_{{\bf n}',{\bf n}}}\bra{\Psi_{\bf n}}
\end{align}
with $d_t=d/(dt)$. Using the explicit form of $\mathcal{L}_0$ we find
\begin{align}
	\frac{\partial\hat{\rho}}{\partial t}	=&\left(-i\frac{E_{\bf n'}-E_{\bf n}}{\hbar}-i\delta\alpha_{{\bf n}',{\bf n}}\hat{a}^\dag\right)C_{{\bf n}',{\bf n}}\ket{{\bf n}',\alpha_{{\bf n}',{\bf n}}}\bra{\Psi_{\bf n}}-\frac{\Gamma}{2}\hat{a}^\dag\alpha_{{\bf n}',{\bf n}} C_{{\bf n}',{\bf n}}\ket{{\bf n}',\alpha_{{\bf n}',{\bf n}}}\bra{\Psi_{\bf n}}\nonumber\\
	&-\frac{\Gamma|\chi|^2}{2}[\Theta_{\bf n}-\Theta_{\bf n'}]^2 C_{{\bf n}',{\bf n}} \ket{{\bf n}',\alpha_{{\bf n}',{\bf n}}}\bra{\Psi_{\bf n}}-\Gamma\chi^*\alpha_{{\bf n}',{\bf n}}[\Theta_{\bf n'}-\Theta_{\bf n}]C_{{\bf n}',{\bf n}}\ket{{\bf n}',\alpha_{{\bf n}',{\bf n}}}\bra{\Psi_{\bf n}}.
\end{align}
Comparing the two results leads to the differential equations
\begin{align}
	d_tC_{{\bf n}',{\bf n}}-\frac{d_t|\alpha|^2}{2}C_{{\bf n}',{\bf n}}=&\left(-i\frac{E_{\bf n'}-E_{\bf n}}{\hbar}-\frac{\Gamma|\chi|^2}{2}[\Theta_{\bf n}-\Theta_{\bf n'}]^2-\Gamma\chi^*\alpha_{{\bf n}',{\bf n}}[\Theta_{\bf n'}-\Theta_{\bf n}]\right) C_{{\bf n}',{\bf n}},\\
	d_t\alpha_{{\bf n}',{\bf n}}=&\left[-i\delta-\frac{\Gamma}{2}\right]\alpha_{{\bf n}',{\bf n}}.
\end{align}
We can now solve for $\alpha_{{\bf n}',{\bf n}}$
\begin{align}
	\alpha_{{\bf n}',{\bf n}}=&\alpha_{{\bf n}',{\bf n}}(0)e^{\left[-i\delta-\frac{\Gamma}{2}\right]t},\\
	|\alpha_{{\bf n}',{\bf n}}|^2=&|\alpha_{{\bf n}',{\bf n}}(0)|^2e^{-\Gamma t},
\end{align}
and use this result in $d_tC_{{\bf n}',{\bf n}}$ together with $\alpha_{{\bf n}',{\bf n}}(0)=\chi(\Theta_{\bf n}-\Theta_{\bf n'})$ to obtain
\begin{align}
	d_tC_{{\bf n}',{\bf n}}=&\left(-i\frac{E_{\bf n'}-E_{\bf n}}{\hbar}-\frac{\Gamma|\chi|^2(1+e^{-\Gamma t})}{2}[\Theta_{\bf n}-\Theta_{\bf n'}]^2+\Gamma|\chi|^2e^{\left[-i\delta-\frac{\Gamma}{2}\right]t}|\Theta_{\bf n'}-\Theta_{\bf n}|^2\right) C_{{\bf n}',{\bf n}}.
\end{align}
The result for $C_{{\bf n}',{\bf n}}$ reads then
\begin{align}
	C_{{\bf n}',{\bf n}}=e^{-i\frac{E_{\bf n'}-E_{\bf n}}{\hbar}t-\frac{\Gamma|\chi|^2}{2}[\Theta_{\bf n}-\Theta_{\bf n'}]^2t-\frac{\Gamma|\chi|^2}{2}[\Theta_{\bf n}-\Theta_{\bf n'}]^2\frac{1-e^{-\Gamma t}}{\Gamma}+\Gamma|\chi|^2|\Theta_{\bf n'}-\Theta_{\bf n}|^2\frac{1-e^{\left[-i\delta-\frac{\Gamma}{2}\right]t}}{i\delta+\frac{\Gamma}{2}}}.
\end{align}
\subsection{Calculation of $O$}
We can now calculate $O$ with
\begin{align}
	O=&-\frac{1}{\hbar^2}\bra{{\bf n},\mathrm{vac}}\hat{D}^\dag\hat{H}_\perp\hat{D}\left\{\int_{0}^{\infty}d\tau e^{\mathcal{Q}\mathcal{L}_0\tau}\left[\sum_{\bf m}P_{\bf m}\hat{D}^\dag\hat{H}_\perp\hat{D}\ket{\Psi_{\bf m}}\bra{\Psi_{\bf m}}\right]\right\}\ket{{\bf n},\mathrm{vac}}+\mathrm{c.c}.\\
	=&-\sum_{\bf n'}\frac{|\bra{\bf n}\hat{H}_\perp\ket{\bf n'}|^2}{\hbar^2}\int_{0}^{\infty} d\tau C_{{\bf n}',{\bf n}}(\tau)\bra{{\bf n}',\chi(\Theta_{\bf n}-\Theta_{\bf n'})}\ket{{\bf n}',\alpha_{{\bf n}',{\bf n}}(\tau)} P_{\bf n}+\mathrm{c.c}
 \end{align}
Using the explicit form of $\alpha_{{\bf n}',{\bf n}}(\tau)$ and the overlap
\begin{align}
	\bra{\chi[\Theta_{\bf n}-\Theta_{\bf n'}]}\ket{\chi[\Theta_{\bf n}-\Theta_{\bf n'}]e^{\left[-i\delta-\frac{\Gamma}{2}\right]t}}=e^{-\frac{|\chi|^2|\Theta_{{\bf n}'}-\Theta_{\bf n}|^2}{2}\left(1+e^{-\Gamma t}-2e^{\left[-i\delta-\frac{\Gamma}{2}\right]t}\right)}
\end{align}
we obtain
\begin{align}
	O=&-\sum_{\bf n'}\frac{2|\bra{\bf n}\hat{H}_\perp\ket{\bf n'}|^2}{\hbar^2}\mathrm{Re}\left[\int_{0}^{\infty} d\tau e^{-i\frac{E_{\bf n'}-E_{\bf n}}{\hbar}t-\frac{\Gamma|\chi|^2}{2}[\Theta_{\bf n}-\Theta_{\bf n'}]^2t-\frac{(i\delta-\frac{\Gamma}{2})}{i\delta+\frac{\Gamma}{2}}|\chi|^2[\Theta_{\bf n}-\Theta_{\bf n'}]^2\left(1-e^{-\left[i\delta+\frac{\Gamma}{2}\right]t}\right)}\right] P_{\bf n}.
\end{align}
\subsection{Calculation of $I$}
We now calculate 
\begin{align}
	I
	=&\frac{1}{\hbar^2}\mathrm{Tr}_{\mathrm{phot}}\left[\bra{{\bf n}}\left\{\int_{0}^{\infty}d\tau e^{\mathcal{Q}\mathcal{L}_0\tau}\left[\sum_{\bf m}P_{\bf m}\hat{D}^\dag\hat{H}_\perp\hat{D}\ket{\Psi_{\bf m}}\bra{\Psi_{\bf m}}\right]\right\}\hat{D}^\dag\hat{H}_\perp\hat{D}\ket{{\bf n}}\right]+\mathrm{c.c}\\
	=&\sum_{\bf n'}\frac{|\bra{\bf n}\hat{H}_\perp\ket{\bf n'}|^2}{\hbar^2}\mathrm{Tr}_\mathrm{phot}\left[\int_{0}^{\infty}d\tau C_{{\bf n},{\bf n}'}(\tau)\ket{\alpha_{{\bf n},{\bf n}'}(\tau)}\bra{\alpha_{{\bf n},{\bf n}'}(0)}\right]P_{\bf n'}.
\end{align}
Here, we have used explicitly that the states $\ket{\mathbf n}$ can be taken outside the photonic trace $\mathrm{Tr}_{\mathrm{phot}}$, since the operator $\ket{\alpha_{\mathbf n,\mathbf n'}(\tau)}\bra{\alpha_{\mathbf n,\mathbf n'}(0)}$ acts solely on the photonic Hilbert space. Note also the interchange of $\mathbf n$ and $\mathbf n'$ compared to the expression for $O$. Performing the photonic trace and evaluating the resulting coherent-state overlap yields
\begin{align}
	I
	=&\frac{1}{\hbar^2}\mathrm{Tr}_{\mathrm{phot}}\left[\bra{{\bf n}}\left\{\int_{0}^{\infty}d\tau e^{\mathcal{Q}\mathcal{L}_0\tau}\left[\sum_{\bf m}P_{\bf m}\hat{D}^\dag\hat{H}_\perp\hat{D}\ket{\Psi_{\bf m}}\bra{\Psi_{\bf m}}\right]\right\}\hat{D}^\dag\hat{H}_\perp\hat{D}\ket{{\bf n}}\right]+\mathrm{c.c}\\
	=&\sum_{\bf n'}\frac{2|\bra{\bf n}\hat{H}_\perp\ket{\bf n'}|^2}{\hbar^2}\mathrm{Re}\left[\int_{0}^{\infty} d\tau e^{-i\frac{E_{\bf n}-E_{\bf n'}}{\hbar}t-\frac{\Gamma|\chi|^2}{2}[\Theta_{\bf n'}-\Theta_{\bf n}]^2t-\frac{(i\delta-\frac{\Gamma}{2})}{i\delta+\frac{\Gamma}{2}}|\chi|^2[\Theta_{\bf n'}-\Theta_{\bf n}]^2\left(1-e^{-\left[i\delta+\frac{\Gamma}{2}\right]t}\right)}\right]P_{\bf n'}.
\end{align}

Together we can then write
\begin{align}
	\frac{dP_{\bf n}}{dt}=O+I=\sum_{{\bf n}'}[-R_{{\bf n}\to{\bf n}'}P_{\bf n}+R_{{\bf n'}\to{\bf n}}P_{{\bf n}'}]
\end{align}
with
\begin{align}
R_{{\bf n}\to{\bf n}'}=&\frac{2|\bra{\bf n}\hat{H}_\perp\ket{\bf n'}|^2}{\hbar^2}\mathrm{Re}\left[\int_{0}^{\infty} dt e^{-i\frac{E_{\bf n'}-E_{\bf n}}{\hbar}t-\frac{\Gamma|\chi|^2}{2}[\Theta_{\bf n}-\Theta_{\bf n'}]^2t-\frac{(i\delta-\frac{\Gamma}{2})}{i\delta+\frac{\Gamma}{2}}|\chi|^2[\Theta_{\bf n}-\Theta_{\bf n'}]^2\left(1-e^{-\left[i\delta+\frac{\Gamma}{2}\right]t}\right)}\right]
\end{align}
which gives the expression shown in the main text when using $\Delta E_{{\bf n'},{\bf n}} =E_{\bf n'}-E_{\bf n}$, $\Delta\Theta_{{\bf n'},{\bf n}}=\Theta_{\bf n}-\Theta_{\bf n'}$, and $(i\delta-\Gamma/2)/(i\delta+\Gamma/2)=\chi/\chi^*$.

\end{widetext}

\section{Influence of the different excitations on the bosonic state}
In this section, we analyze the influence of the different onsite excitations on the state of the system. To do this, we examine for the bosonic system studied in the main text, the effect of different on-site bosonic occupation cutoffs $N_{\mathrm{max}}$. 

\begin{figure}[!hbtp]
	\centering
	\includegraphics[width=0.48\textwidth]{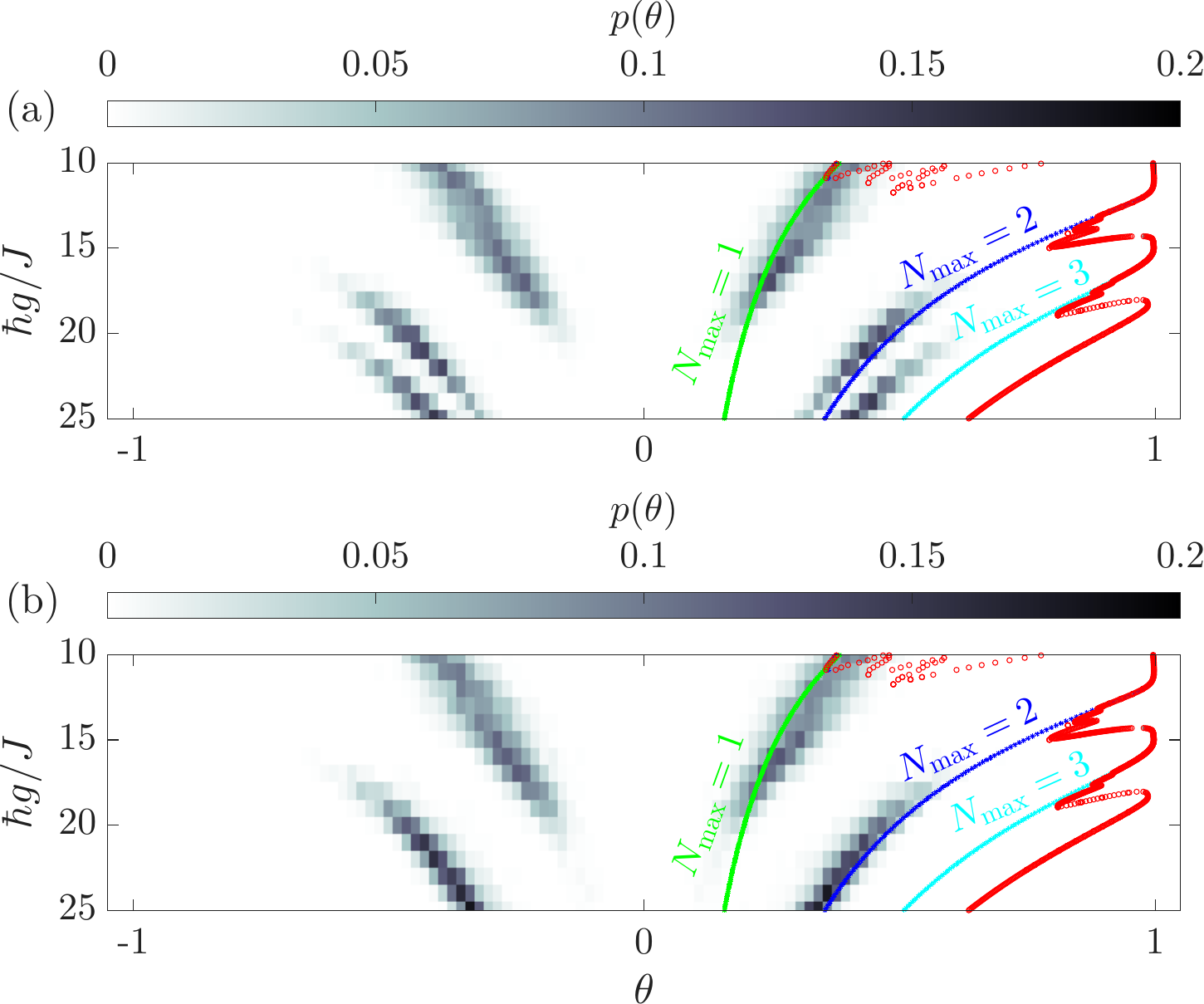}
	\caption{Results for $p(\theta)$ as a function of $g$ and $\theta$ obtained with the DSRE method are shown as shaded regions for the onsite boson-number cutoff (a) $N_{\mathrm{max}}=N$ and (b) $N_{\mathrm{max}}=2$. The DSRE results shown in (a) are identical to those presented in Fig.~1(f) of the main text. The parameters are $U_0 = 40J$, $\hbar\delta = 5J$, $\hbar\Gamma = 3J$, $L = 224$, and half filling, $N=L/2$. The colored lines show $|\theta_{\mathrm{th}}|$ obtained using the MF+fluct. method for a system size of $L=8$. The green, blue, and cyan curves correspond to onsite boson-number cutoffs of $N_{\mathrm{max}}=1$, $N_{\mathrm{max}}=2$, and $N_{\mathrm{max}}=3$, respectively.}
	\label{fig:S1}
\end{figure}

In Fig.~\ref{fig:S1}(a), we show the distribution $p(\theta)$ obtained from the DSRE method at time $Jt/\hbar = 1000$ as the shaded region, with the MF+fluct. predictions for $\theta_{\mathrm{th}}$ computed using different on-site occupation cutoffs: $N_{\mathrm{max}}=1$ (green), $N_{\mathrm{max}}=2$ (blue), $N_{\mathrm{max}}=3$ (cyan), and no cutoff (red), for a system of size $L=8$. The DSRE results and the MF+fluct. results for $N_{\mathrm{max}}=1$ and for the case without a cutoff are identical to those shown in Fig.~1(f) of the main text.

As shown by the MF+fluct. results in Fig.~\ref{fig:S1}(a), the bistability regions depend on the local Hilbert-space cutoff $N_{\mathrm{max}}$. Specifically, no bistability is observed for $N_{\mathrm{max}}=1$, a single bistability region emerges for $N_{\mathrm{max}}=2$, and two distinct bistability regions appear for $N_{\mathrm{max}}=3$. These results demonstrate that the local Hilbert-space structure has a pronounced quantitative influence on both the location and the number of bistability regions.

Comparing the DSRE and MF+fluct.~methods, we find that for the smallest values of $g$ considered, the first peak of $p(\theta)$ obtained from the DSRE agrees very well with the MF+fluct. result for $N_{\mathrm{max}}=1$ (green). This agreement persists as $g$ increases until the DSRE predicts the onset of a bistable region, before the distribution fully transitions to a larger value of $\theta$ at approximately $\hbar g/J \approx 19$. Beyond this point, the DSRE results agree well with the MF+fluct. results obtained for $N_{\mathrm{max}}=2$ (blue). For even larger values of $g$, the DSRE predicts a second bistable region, where an additional peak in $p(\theta)$ emerges at larger values of $\theta$. This peak agrees best with the MF+fluct. results for $N_{\mathrm{max}}=3$ (cyan) compared to the red, blue, and green MF+fluct. branches. However, the correspondence is less quantitative than the one observed for lower $g$ values.

To further demonstrate that the local Hilbert-space structure determines the bistability regions, we perform additional DSRE simulations with the local Hilbert-space cutoff fixed to $N_{\mathrm{max}}=2$, shown as the shaded region in Fig.~\ref{fig:S1}(b). These results are again compared with the MF+fluct. calculations for different values of $N_{\mathrm{max}}$.

We find that the third $\theta$ branch visible in Fig.~\ref{fig:S1}(a) disappears once the cutoff $N_{\mathrm{max}}=2$ is imposed. More specifically, the shaded region below the cyan curve, which is clearly present in Fig.~\ref{fig:S1}(a), is absent in Fig.~\ref{fig:S1}(b). This provides strong evidence that the additional bistability branch originates from the enlarged local Hilbert space and directly links the emergence of bistability regions to the local Hilbert-space structure.

We now comment on the discrepancies observed between the DSRE and MF+fluct. approaches. One possible source is that the DSRE simulations are necessarily performed over a finite evolution time and may therefore not have fully converged to the stationary state. Another possible origin is the thermal-state assumption underlying the MF+fluct. approach. While the DSRE method allows for non-thermal stationary states, the MF+fluct. approach assumes explicitly that the bosons are in a thermal state. Overall, we find good qualitative agreement, together with partial quantitative agreement, between the two methods. Notably, the DSRE simulations consistently favor states with lower bosonic occupancies than those predicted by the MF+fluct. approach.

\begin{figure}[!hbtp]
	\centering
	\includegraphics[width=0.48\textwidth]{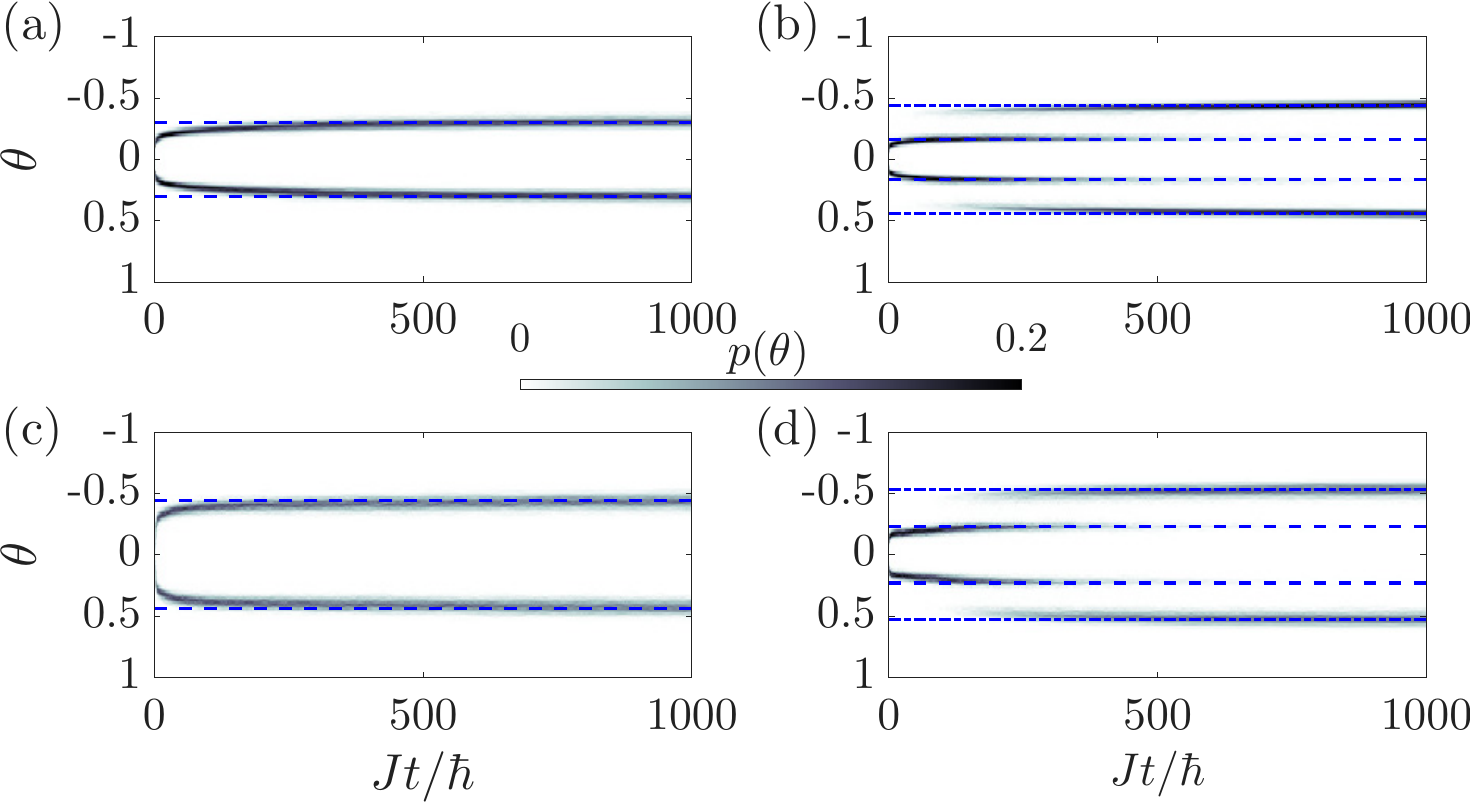}
	\caption{(a)-(d) Time evolution of $p(\theta)$ for spins (a),(b) and Fermions (c), (d) obtained from the DSRE method. The parameters are (a) $\hbar g/J=25$, (b) $\hbar g/J=35$, (c) $\hbar g/J=8$, and (d) $\hbar g/J=13$. The remaining parameters for (a) and (b): $U_0 = 30J$, $\hbar\delta = 10J$, $\hbar \Gamma = 3J$, $L = 224$, total magnetization $\sum_j\hat{S}^z_j=-L/2$; for (c) and (d): $U_0 = 20J$, $\hbar\delta = 5J$, $\hbar\Gamma = 3J$, $L = 224$, with quarter filling $N_\downarrow = L/4$, $N_\uparrow = L/4$. Guide to the eye: the blue dashed line indicates the position of the first $\theta$-peak, while the blue dash-dotted line indicates the position of the second $\theta$-peak.}
	\label{fig:S2}
\end{figure}

\section{Relaxation Dynamics of Spins and Fermions}

In this section, we present the dynamics of spins and fermions, analogous to the bosonic dynamics shown in Fig.~2 of the main text. We show that the general behavior is very similar for all three models. 

For each case, we compute $p(\theta)$ using the DSRE method over a finite evolution time $Jt/\hbar$. The resulting peak of the distribution, shown as the shaded region in Fig.~\ref{fig:S2}, is used to track the evolution of $\theta$. We consider values of $\hbar g/J$ for both spins [(a) and (b)] and fermions [(c) and (d)] that illustrate two distinct relaxation behaviors: direct relaxation to a finite value of $\theta$ [(a) and (c)] and relaxation through an intermediate metastable value before converging to the final steady-state value [(b) and (d)].

For the spin-1 system, we choose (a) $\hbar g/J=25$ and (b) $\hbar g/J=35$, corresponding to values slightly below and above the bistable region shown in Fig.~1(b) of the main text. In Fig.~\ref{fig:S2}(a), the distribution relaxes directly to a value of $|\theta|\lesssim 0.5$. In contrast, Fig.~\ref{fig:S2}(b) shows that the distribution initially peaks around $|\theta|\lesssim 0.2$, before a second peak emerges near $|\theta|\lesssim 0.5$. As time evolves, the probability gradually transfers to the larger-$|\theta|$ peak, and the system ultimately relaxes to this state, similarly to the bosonic case shown in Fig.~2 of the main text.

A similar behavior is found for the fermionic system. We consider (c) $\hbar g/J=8$ and (d) $\hbar g/J=13$, corresponding to values slightly below and above the bistable region shown in Fig.~1(d) of the main text. Below the bistable region, the distribution relaxes directly to the stationary state [Fig.~\ref{fig:S2}(c)]. Above the bistable region, the system remains in a metastable configuration for a transient period before eventually relaxing to its final stationary state [Fig.~\ref{fig:S2}(d)].

These results demonstrate that the dynamical behavior highlighted in the main text for bosons is not unique to that system, but is also observed for spins and fermions.

\section{Time-dependent matrix product states for hybrid quantum systems}

In this section, we briefly describe the time-dependent matrix product state (tMPS) implementation used to simulate the full quantum dynamics of the matter-cavity Lindblad master equation [Eq.~(1) in the main text]. We also present the time-dependent evolution corresponding to the long-time results shown in Fig.~3(f) of the main text. The tMPS results are obtained for interacting bosonic atoms  with a local Hilbert space of at most two bosons per site coupled to the dissipative cavity field (see main text).

\subsection{Method description}

The numerically exact simulations for the time evolution of the atom-cavity density matrix with the Lindblad master equation for the case of a one-dimensional Bose-Hubbard model coupled to a dissipative cavity have been performed by employing a recent implementation of a matrix product states (MPS) method  \cite{HalatiKollath2020, HalatiKollath2020b, TolleHalati2026b}.
To deal with the cavity losses we employ the stochastic unravelling of the master equation with quantum trajectories \cite{DalibardMolmer1992, GardinerZoller1992, Daley2014}.
The resulting effective Hamiltonian contains both short-range atomic terms and the global-range coupling to the cavity. To overcome this we make use of a variant of the quasi-exact time-dependent variational matrix product state (tMPS) based on both the Trotter-Suzuki decomposition of the time evolution propagator \cite{WhiteFeiguin2004, DaleyVidal2004, Schollwoeck2011} and the time-dependent variational principle approach (TDVP) \cite{HaegemanVerstraete2011, HaegemanVerstraete2016}.
In this tMPS+TDVP implementation, the evolution under the local Hamiltonian is performed with two-site gates stemming from the Trotter-Suzuki decomposition, while the coupling to the cavity is treated within the TDVP approach \cite{TolleHalati2026b}.
The method has been implemented by employing the ITensor Library \cite{FishmanStoudenmire2020}. 
Additional details regarding the implementation, benchmarks and comparisons with implementations based solely on TDVP, or employing swap gates, can be found in Refs.~\cite{HalatiKollath2020b, TolleHalati2026b}.

The convergence of the results presented in Fig.~3(f) in the main text was ensured by the following convergence parameters: a maximal bond dimension of $150$ states for $L=14$ (for $L=16$ results shown in the next section the bond dimension is of $200$ states), which ensured a truncation error of at most $5\times10^{-9}$ throughout the evolution, a time-step of $dt J/\hbar=4\times10^{-3}$, the local Hilbert space of the bosonic atoms contains maximum two bosons per site, $N_{\mathrm{max}}=2$,
and the adaptive cutoff of the local Hilbert space of the photonic mode ranged between $N_\text{pho}=26$ and $N_\text{pho}=10$. 
The results are averaged over at least $475$ quantum trajectories.

\subsection{Dissipative dynamics results}

\begin{figure}[!hbtp]
	\centering
	\includegraphics[width=0.4\textwidth]{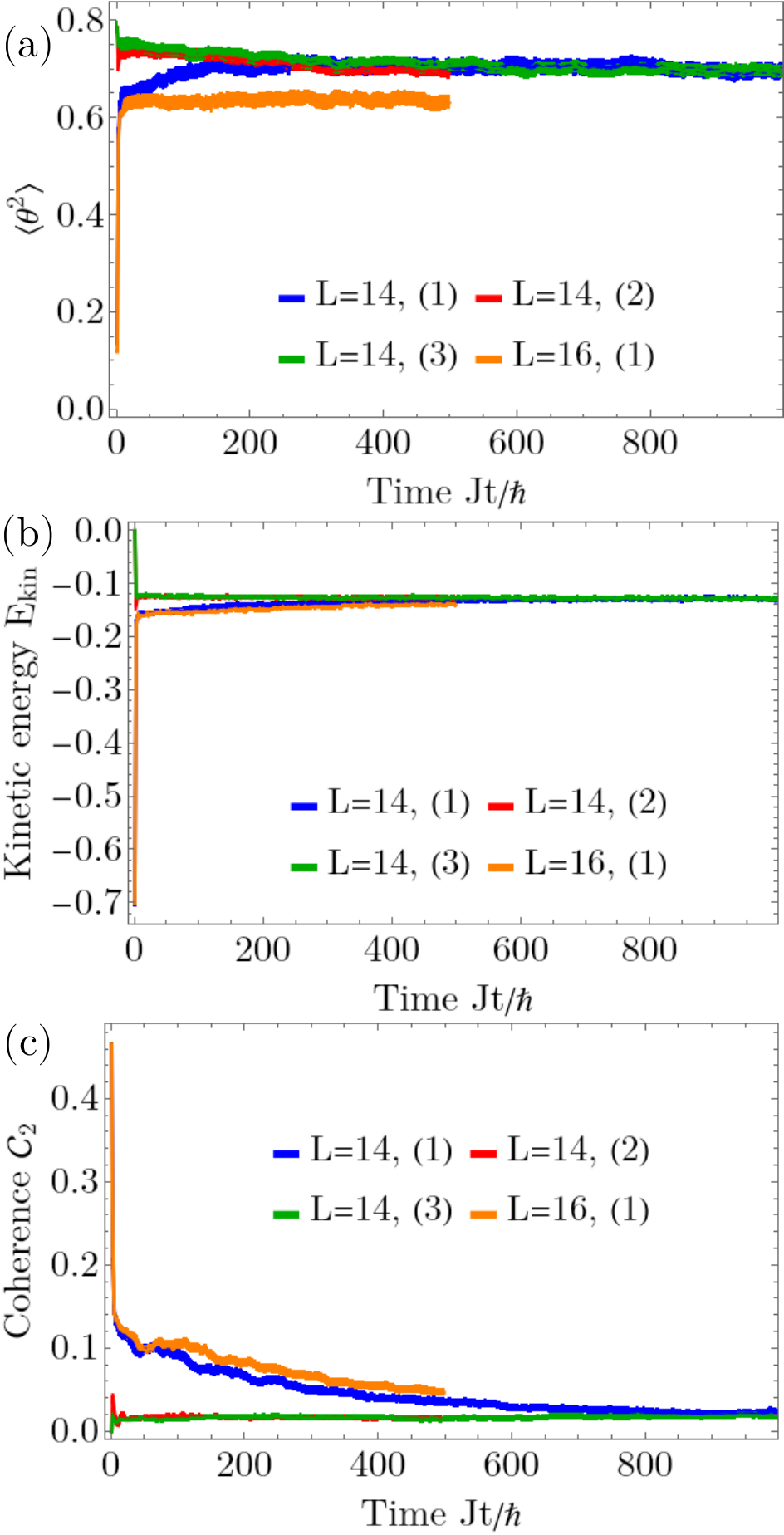}
	\caption{tMPS time evolution of several observables: (a) squared imbalance $\langle\theta^2\rangle$, (b) kinetic energy $E_\text{kin}$, (c) coherence over two sites $\mathcal{C}_2$, [see Eq.~\eqref{eq:observables} for definitions]. The results are obtain for bosonic atoms with $N_\text{max}=2$ for fixed density $N=L/2$, with $L=14$ and $L=16$, and the following parameters $\hbar g/J =10$, $U_0/J=40$, $\hbar \delta/J=10$, $\hbar\Gamma/J=3$. 
	We consider several initial states: (1) ground state of the Bose-Hubbard model for $U_0/J=40$ and empty cavity (blue and orange curves); (2) mixture of Fock states sampled from the DSRE steady state density matrix and empty cavity (red curves); (3) mixture of Fock states sampled from the DSRE steady state density matrix and the associated cavity coherent states for each imbalance (green curves).
	All the curves contain error bars quantifying the stochastic uncertainty stemming from the sampling of the quantum trajectories.}
	\label{fig:MPS1}
\end{figure}

\begin{figure}[!hbtp]
	\centering
	\includegraphics[width=0.4\textwidth]{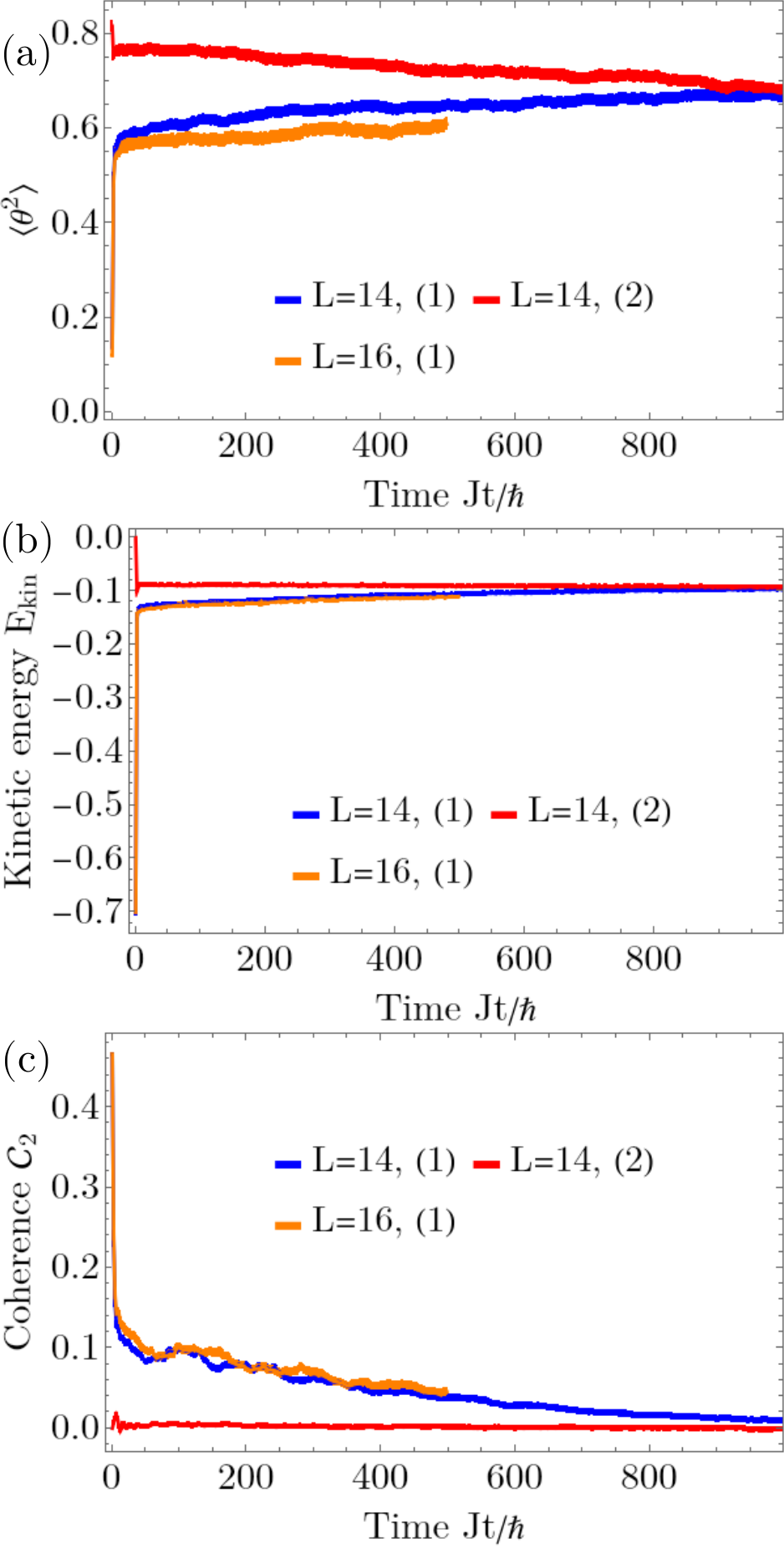}
	\caption{tMPS time evolution of several observables: (a) squared imbalance $\langle\theta^2\rangle$, (b) kinetic energy $E_\text{kin}$, (c) coherence over two sites $\mathcal{C}_2$, [see Eq.~\eqref{eq:observables} for definitions]. The results are obtain for bosonic atoms with $N_\text{max}=2$ for a fixed density $N=L/2$, with $L=14$ and $L=16$, and the following parameters $\hbar g/J =12$, $U_0/J=40$, $\hbar \delta/J=10$, $\hbar\Gamma/J=3$. 
	We consider two initial states: (1) ground state of the Bose-Hubbard model for $U_0/J=40$ and empty cavity (blue and orange curves); (2) mixture of Fock states sampled from the DSRE steady state density matrix and empty cavity (red curves).
	All the curves contain error bars quantifying the stochastic uncertainty stemming from the sampling of the quantum trajectories.}
	\label{fig:MPS2}
\end{figure}

In the following, we supplement the numerical tMPS results of the coupled atom-cavity system shown for long times in Fig.~3(f) with the corresponding time-dependent results towards the long time state.
We focus on two values of the coupling strength, $\hbar g=10J$ in Fig.~\ref{fig:MPS1}, before the fluctuation-induced bistability, and $\hbar g=12J$ in Fig.~\ref{fig:MPS2}, close to the fluctuation-induced bistability.
We show the time dependence of the density imbalance squared, $\langle \theta^2 \rangle$, kinetic energy, $E_\text{kin}$, and coherence over two sites, $\mathcal{C}_2$, 
\begin{align}
	\label{eq:observables}
\langle \theta^2 \rangle &= \frac{\langle \hat{\Theta}^2 \rangle}{\Theta_\text{max}^2}= \frac{4}{L^2}\left\langle \left(\sum_{j=1}^{L}(-1)^j\hat{n}_j \right)^2 \right\rangle, \\
E_\text{kin}&=-J\frac{1}{L-1}\sum_{j=1}^{L-1} \left\langle (b^\dagger_j b_{j+1}+b^\dagger_{j+1} b_{j}) \right\rangle, \nonumber\\
\mathcal{C}_2&=\frac{1}{L-2}\sum_{j=1}^{L-2} \left\langle (b^\dagger_j b_{j+2}+b^\dagger_{j+2} b_{j}) \right\rangle. \nonumber
\end{align}
In order to understand how close the numerical results at the final simulation are to the steady state we consider different initial states for $L=14$,  {\it (1)} the Bose-Hubbard ground state for the atoms and vacuum state for the cavity (blue curves in Fig.~\ref{fig:MPS1} and Fig.~\ref{fig:MPS2}), {\it (2)} a mixed state consisting of atomic Fock states sampled from the DSRE steady state and vacuum state for the cavity (red curves in Fig.~\ref{fig:MPS1} and Fig.~\ref{fig:MPS2}), and {\it (3)} a mixed state consisting of atomic Fock states sampled from the DSRE steady state and the corresponding coherent states resulting from the polaron transformation (green curves in Fig.~\ref{fig:MPS1}).

In Fig.~\ref{fig:MPS1}, for $\hbar g=10J$, we observe for $L=14$ that the density imbalance $\langle \theta^2 \rangle$ exhibits a quick evolution away from the initial states, followed by a slower evolution towards the steady-state value. In particular, for times of $tJ/\hbar\approx 150$, the evolution from all three considered steady states converges to the same value, which seems to remain steady within the stochastic uncertainty of the quantum trajectories [Fig.~\ref{fig:MPS1}(a)].
However, finite-size effects seem important in this regime for the value of $\langle \theta^2 \rangle$, since deviations between blue curve for $L=14$ in comparison to orange curve for $L=16$ in Fig.~\ref{fig:MPS1} are seen.
Interestingly, the coherences between the neighboring sites, quantified by the kinetic energy $E_\text{kin}$, and over two sites, given by $\mathcal{C}_2$, are initially rapidly suppressed towards the long-time value when starting from the ground state. Also for the case {\it (2)} when starting from incoherent Fock states the coherences are rapidly generated approaching the same long-time value.
In all cases they stabilize to finite values on times longer than in the case of the imbalance, $tJ/\hbar\approx 300$ for $E_\text{kin}$ [Fig.~\ref{fig:MPS1}(b)] and $tJ/\hbar\approx 800$ for $\mathcal{C}_2$ [Fig.~\ref{fig:MPS1}(c)]. 
For $E_\text{kin}$ and $\mathcal{C}_2$ the results for $L=14$ and  $L=16$ seem to be consistent within the stochastic uncertainty.
The final coherences show the importance of the off-diagonal matrix elements of the atomic density matrix for the tMPS results. The presence of these sizable coherences can explain the deviations seen between the MPS results and the DSRE results.

For larger values of the coupling, $\hbar g=12J$, shown in Fig.~\ref{fig:MPS2}, close to the regime identified for the fluctuation-induced bistability (see main text), the time needed to reach the steady-state values increases considerably. This is consistent with the dynamical DSRE results discussed in the main text.
We observe that the values of $\langle \theta^2 \rangle$ for the two considered initial states start agreeing only close to the final evolution times considered $tJ/\hbar\approx 1000$. 
In particular, this time is almost an order of magnitude longer than for $\hbar g=10J$ shown in Fig.~\ref{fig:MPS1}(a).
We note that we cannot fully exclude that an even slower evolution might show up at later times. 
Interestingly, in the bistability regime the imbalance dynamics seems to have comparable, or even longer, timescales than the coherences as shown for $E_\text{kin}$ [Fig.~\ref{fig:MPS2}(b)] and $\mathcal{C}_2$ [Fig.~\ref{fig:MPS2}(c)]. 
While the kinetic energy evolves towards a finite value at long times, it is not clear if the coherence $\mathcal{C}_2$ survives in the steady state, however it maintains a finite value for a large part of the dynamics.
Thus, even for relatively small system sizes, the bistable regime is characterized by very long timescales, indicating the presence of bistability.

\end{document}